\documentclass[aps,prl,reprint,floatfix,superscriptaddress]{revtex4-1}

\pdfoutput=1
\usepackage{graphicx}
\usepackage{color}
\usepackage{natbib}
\usepackage{latexsym}
\usepackage{siunitx}
\usepackage{amsmath}

\begin{document}

\title{\LARGE Determining interface dielectric losses in superconducting coplanar waveguide resonators}

\author{W. Woods}
\thanks{wayne.woods@ll.mit.edu; These authors contributed equally.}
\affiliation{MIT Lincoln Laboratory, 244 Wood Street, Lexington, MA 02421, USA}
\author{G. Calusine}
\thanks{wayne.woods@ll.mit.edu; These authors contributed equally.}
\affiliation{MIT Lincoln Laboratory, 244 Wood Street, Lexington, MA 02421, USA}
\author{A. Melville}
\thanks{wayne.woods@ll.mit.edu; These authors contributed equally.}
\affiliation{MIT Lincoln Laboratory, 244 Wood Street, Lexington, MA 02421, USA}
\author{A. Sevi}
\affiliation{MIT Lincoln Laboratory, 244 Wood Street, Lexington, MA 02421, USA}
\author{E. Golden}
\affiliation{MIT Lincoln Laboratory, 244 Wood Street, Lexington, MA 02421, USA}
\author{D. K. Kim}
\affiliation{MIT Lincoln Laboratory, 244 Wood Street, Lexington, MA 02421, USA}
\author{D. Rosenberg}
\affiliation{MIT Lincoln Laboratory, 244 Wood Street, Lexington, MA 02421, USA}
\author{J. L. Yoder}
\affiliation{MIT Lincoln Laboratory, 244 Wood Street, Lexington, MA 02421, USA}
\author{W. D. Oliver}
\affiliation{MIT Lincoln Laboratory, 244 Wood Street, Lexington, MA 02421, USA}
\affiliation{Research Laboratory of Electronics, Massachusetts Institute of Technology, Cambridge, MA 02139, USA}
\affiliation{Department of Physics, Massachusetts Institute of Technology, Cambridge, MA 02139, USA}

\date{\today}

\begin{abstract}
Superconducting quantum computing architectures comprise resonators and qubits that experience energy loss due to two-level systems (TLS) in bulk and interfacial dielectrics. Understanding these losses is critical to improving performance in superconducting circuits. In this work, we present a method for quantifying the TLS losses of different bulk and interfacial dielectrics present in superconducting coplanar waveguide (CPW) resonators. By combining statistical characterization of sets of specifically designed CPW resonators on isotropically etched silicon substrates with detailed electromagnetic modeling, we determine the separate loss contributions from individual material interfaces and bulk dielectrics. This technique for analyzing interfacial TLS losses can be used to guide targeted improvements to qubits, resonators, and their superconducting fabrication processes.
\end{abstract}

\maketitle

\section{Introduction}
Two-level systems (TLS) have been identified as critical contibutors limiting performance in superconducting qubits and resonators.\cite{Oliver2013,Martinis2005,Gao2008a,Sage2011}  While the microscopic origin of many TLS are unknown, it is well established that ensembles of TLS contribute to energy loss in superconducting devices through their interaction with the electric fields present in the bulk dielectric materials and interfaces.  Efforts to mitigate these losses have employed techniques from materials science, fabrication process engineering, and microwave device design.  Materials improvements have focused on lowering TLS defect densities in bulk materials \cite{Paik2010} or removing TLS-containing dielectrics \cite{OConnell2008}.  Fabrication process advancements have included steps aimed at reducing TLS loss through substrate preparation \cite{Megrant2012, Bruno2015} and chemical residue removal.\cite{Quintana2014} Design changes to qubits and resonators have reduced or shifted the electromagnetic (EM) fields interacting with surrounding material interfaces with the goal of minimizing TLS losses.\cite{Paik2011,Khalil2011,Wang2015,Chu2016,Dial2016,Gambetta2017, Calusine2018, Barends2010} Collectively, these advances have led to qubit $T_1$ times exceeding 50 $\mu$s \cite{Jin2015,Yan2016,Rigetti2012} and planar superconducting coplanar waveguide (CPW) resonators quality factors ($Q_i$) in excess of 2 million at single-photon energy levels.\cite{Calusine2018} \par

Further progress in reducing TLS losses has been hindered by an inability to isolate the contributions from separate sources of TLS loss.  Several previous efforts to quantify interface losses have attemped to shift and reduce surface participation by using anisotropic substrate trenching. While this did reduce overall TLS loss and thereby improve $T_1$ and $Q_i$,\cite{Vissers2012a,Bruno2015,Gambetta2017, Calusine2018} anisotropic trenching reduces multiple sources of TLS loss by a similar amount, such that their relative contributions to total resonator loss are largely unknown.  In contrast, EM simulations showed that the losses associated with certain TLS-containing regions could be separately accentuated or suppressed through the use of isotropic etching.\cite{Chu2016}  Neverthless, previous work has only studied changes in aggregate TLS loss or put bounds on individual interface losses and it has not been possible to distinguish and quantify loss contributions from individual dielectrics and material interfaces.  Other efforts have succeeded in characterizing the loss contributions from bulk deposited dielectrics \cite{Martinis2005, Weber2011} and superconducting metals \cite{Minev2013}, but these devices differ significantly from the modern planar circuits that are of interest for superconducting quantum computing circuit architectures.  

In this work, we use statistical characterization of sets of four different CPW resonator designs with isotropically etched substrates, combined with detailed EM modeling, to determine the individual contributions to aggregate TLS losses of multiple material interface dielectrics and the bulk silicon substrate.  We then perform additional characterization of a series of devices with widely varying losses and EM participation ratios in order to verify a participation ratio-based loss model.  Furthermore, we perform simulated experiments to compare this approach to previous approaches using anisotropically etched devices and to quantify the measurement resources required to isolate individual interface losses within a certain tolerance.  These results indicate that this technique can provide critical insight into the sources of loss in modern superconducting quantum circuits. 
\par

\section{Results}

\begin{figure}
\includegraphics[scale=1]{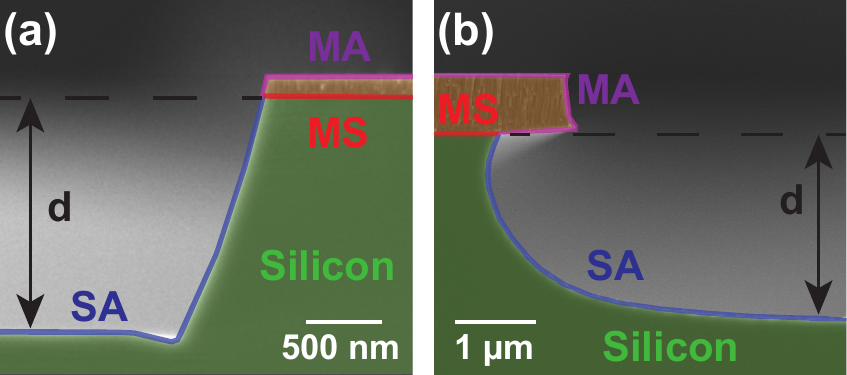}
\caption{\label{fig1} Cross-sectional SEM images of (a) an anisotropically etched TiN resonator, and (b) an isotropically etched TiN resonator.  Dielectric regions are false-colored: metal-to-substrate interface (MS, red), substrate-to-air/vacuum interface (SA, blue), metal-to-air/vacuum interface (MA, purple), and bulk silicon substrate (Si, green).  The trench depth is \textit{d}.   }
\end{figure}

TLS-related losses in superconducting CPW resonators can be analyzed by applying an EM participation ratio model similar to those used in Refs. \cite{Wenner2011,Wang2015,Dial2016,Gambetta2017, Calusine2018}.   In this model, the losses associated with TLS in a given device are a linear combination of the product of the loss tangents ($\tan{\delta_i}$) associated with each dielectric region $i$, and  the participation ratio $p_i$, the fraction of the total electric field energy stored in that region. Each interface contains an unknown combination of dielectrics and fabrication residues, and in our analysis, we assign a unique $\tan{\delta_i}$ to each interface that is exposed to a distinct fabrication process.  We represent our device interfaces using four dielectric regions, as shown in the cross-section in Fig. 1: metal-to-substrate (MS, red), substrate-to-air/vacuum (SA, blue), metal-to-air/vacuum (MA, purple), and the bulk silicon (Si, green).  These regions are generally expected to have distinct dielectric values, thicknesses, and loss tangents.  In this work, we alternatively describe device region losses using scaled participation ratios $P_i$ and ``loss factors'' $x_i$ which are normalized by the dielectric constant and thickness of the defect regions, as detailed in \cite{Calusine2018, Supplemental} and represented in equation 1: 
\begin{equation}
\frac{1}{Q_{TLS}}=\sum\limits_{i}{p_i\tan{\delta_i}}= \sum\limits_{i}{P_ix_i}
\end{equation}
We can generate a matrix representation of Eq. 1 for multiple distinct device geometries sharing a common set of interface properties.  For this case, a column vector of device inverse $Q_{TLS}$'s is determined by a participation matrix $\boldsymbol{P}$ consisting of rows of device participation ratios multiplied by a loss factor column vector $\vec{x}$.  The device $Q_{TLS}$'s are derived from measurement, while the participation ratios of the dielectric regions in our devices are determined from two-dimensional electrostatic simulations using COMSOL \cite{COMSOL}. With these values, we extract the loss factors in Eq. 1 using a linear least-squares fit contrained such that each $x_i$ $>$ 0. \par

We previously used the EM participation model (Eq. 1) to analyze aggregate TLS losses in CPW resonators with anisotropically etched silicon substrates \cite{Calusine2018} such as the one shown in Fig. 1(a).  Anisotropic trenching leads to lower overall surface participation, but the participation ratios of each of the dielectric regions is reduced by a similar amount as the trench depth \textit{d} is increased. This nearly proportional scaling of the participation ratios results in an ill-conditioned participation matrix used to solve Eq. 1.  As a result, generated loss factor solutions are highly sensitive to variations in the input $Q_{TLS}$, leading to large uncertainty when estimating the losses associated with an individual dielectric region.  Accordingly, the precise assignment of loss factor values to individual interfaces is practically unfeasible.  While it is technically possible to reduce the solution uncertainty by reducing the error in the estimates of $Q_{TLS}$, this improves only as $\sqrt{N}$, where $N$ is the number of devices measured, resulting in prohibitively large measurement resource requirements.  In this work, we instead focus on creating a better conditioned participation matrix through the use of isotropically etched CPW resonators. \par

An example cross-section of an isotropically etched TiN resonator is shown in Fig. 1(b).  The false-colored scanning electron microscope (SEM) image also depicts the four dielectric regions that we analyzed: MS, SA, MA, and Si. The simulated participation ratio vectors associated with isotropically etched resonators indicate that these vectors scale with device geometry (center trace width \textit{w}, gap to ground \textit{g}, and etch depth \textit{d}) less proportionally than anisotropically trenched devices of comparable size.\cite{Supplemental, Chu2016, Wenner2011, Gambetta2017}  This non-proportional scaling has a significant impact on the metrics that quantify the singularity of the participation matrix, in particular the condition number $\kappa(\boldsymbol{P})$.  The condition number $\kappa(\boldsymbol{P})$ of the participation matrix $\boldsymbol{P}$ in Eq. 1 relates the uncertainty in the mean $Q^{-1}$ values to the uncertainty in the extracted loss factors.  For the case of four geometries and four loss factors, the ideal participation matrix is a 4x4 identity matrix with condition number = 1. In this case, the uncertainty in the extracted loss factors is equal to the uncertainty of the mean measured $Q^{-1}$ values in Eq. 1. In general, participation matrices with larger condition numbers generate solutions with greater uncertainty than would be determined solely by measurement statistical variance.\cite{Aster2016}  \par

\begin{figure}
\includegraphics[scale=1]{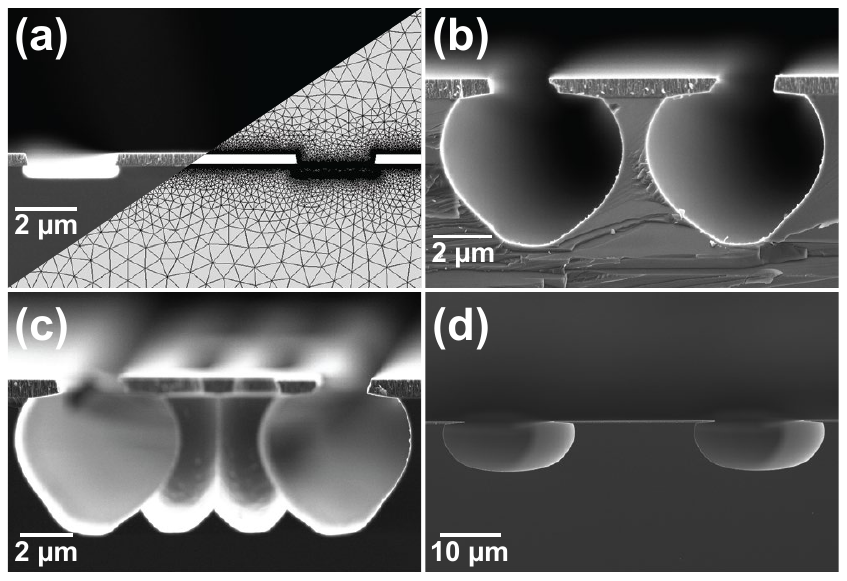}
\caption{\label{fig2} Cross-sectional SEM images of isotropically etched TiN CPW resonators (a) MS-heavy resonator (\textit{w}, \textit{g}, \textit{d}) = (6 $\mu$m, 3 $\mu$m, 0.28 $\mu$m) and the corresponding COMSOL simulation mesh.  (b) SA-heavy resonator, (\textit{w}, \textit{g}, \textit{d}) = (6 $\mu$m, 1 $\mu$m, 4.5 $\mu$m).  (c) MA-heavy partially suspended resonator, (\textit{w}, \textit{g}, \textit{d}) = (8 $\mu$m, 1 $\mu$m, 4.5 $\mu$m). (d) Si-heavy resonator, (\textit{w}, \textit{g}, \textit{d}) = (28 $\mu$m, 14 $\mu$m, 10.9 $\mu$m). }
\end{figure}

For the case of trenched CPW resonators, we determined that isotropic trenching greatly reduces our participation matrix condition number $\kappa(\boldsymbol{P})$ as compared to the case of anisotropically trenched devices.  We performed a constrained search over the range of geometries accessible to our isotropic etch fabrication process (trench depth \textit{d} $\leq$ 11 $\mu$m) in order to determine a set of four structures that minimize $\kappa(\boldsymbol{P})$.  The cross-section of these four CPWs are shown in Fig. 2 (a-d).  Fig. 2(a) shows the CPW cross-section designed to be `MS-heavy' because it maximizes the MS interface region participation relative to the other regions. The shallow trenching in this geometry forms an essentially planar structure comparable to untrenched planar qubits\cite{Barends2013} and CPW resonators.\cite{Megrant2012}  The CPW cross-sections shown in Fig. 2(b) and Fig. 2(d) rely on deep trenching to achieve `SA-heavy' and `Si-heavy' structures, respectively, by varying the center trace dimension and gap-to-ground spacing. The `MA-heavy' cross-section shown in Fig. 2(c) differs compared to the others shown in Fig. 2 in that its signal line is completely undercut for a significant fraction ($\sim$ 85\%) of the total resonator length, and the suspended structure is supported with periodically placed Si posts. These suspended CPWs shift a greater fraction of the total participation to the MA interface than is possible with anisotropic or isotropic etching alone.  See \cite{Supplemental} for details.  The condition number $\kappa(\boldsymbol{P})$ of the participation matrix generated by these four isotropically etched CPW resonator geometries is approximately a factor of 55 lower that what was previously possible using anisotropically trenched devices.

The isotropically etched CPW devices shown in Fig. 1 and Fig. 2 were fabricated using a subtractive etch process on high-resistivity ($\geq$ 3500 $\Omega$-cm) 200mm (001) silicon substrates, similar to the process described in \cite{Calusine2018}. In this work, however, we used a metal thickness of 450 nm or 750 nm, and adjusted the total chlorine-based etch time such that, regardless of the TiN thickness, the substrate was minimally etched. Then, instead of immediately stripping the photoresist, we rinsed the wafer in deionized water and subjected it to a second, fluorine-based plasma etch to isotropically etch the underlying silicon substrate.  The total isotropic etch time was adjusted to control the trench depth \textit{d} and the amount of undercutting. The remaining photoresist was then stripped.  Aside from varying the etch time and using two metal thicknesses, we used a nominally identical fabrication process for all devices. After fabrication, each device geometry was characterized using cross-sectional SEM to refine the simulation geometry input into COMSOL. This improves the physical accuracy of the participation matrix $\boldsymbol{P}$ in Eq. 1. The cross-section in Fig. 2 (a) also shows the COMSOL mesh used for the CPW in the right-half of the figure.  \par

\begin{figure*}
\includegraphics[scale=1]{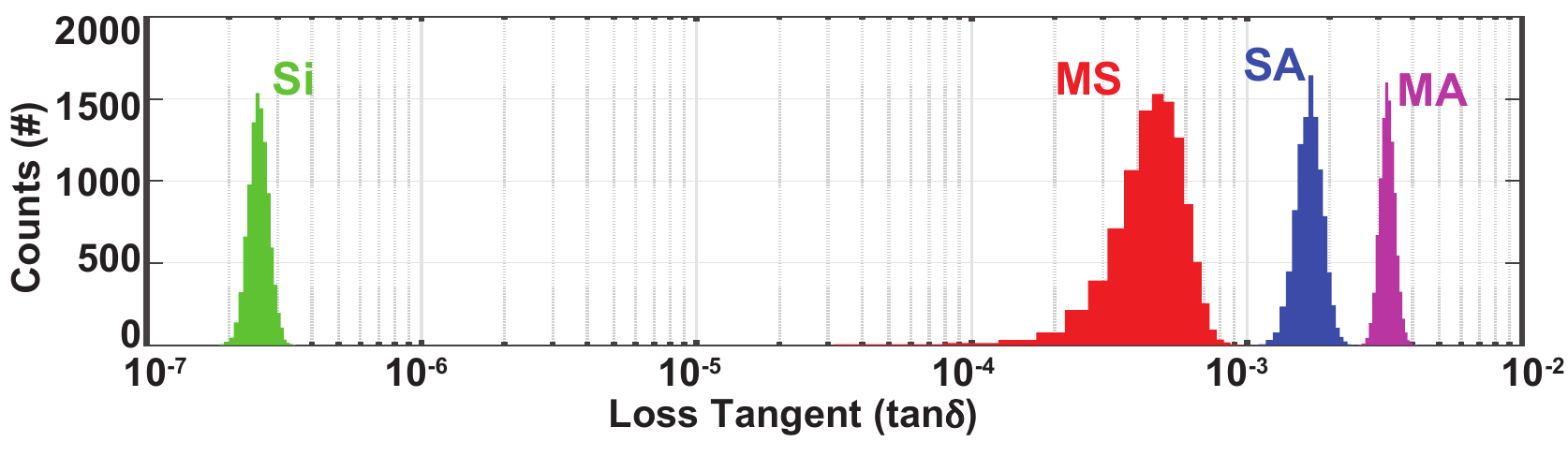}
\caption{\label{fig3} Extracted loss tangents of the metal-to-substrate (MS), substrate-to-air/vacuum (SA), metal-to-air/vacuum (MA), and the silicon substrate (Si) dielectric regions.}
\end{figure*}

In order to account for device-to-device variation and generate the statistics necessary to estimate the uncertainty in the extracted interface losses, we characterized many nominally identical copies of each resonator.   This was achieved by frequency-multiplexing five resonators in the 5 to 6 GHz range with the same width, gap, and trench depth on 5 mm x 5 mm chips and measuring many such identical chips within individual microwave-connectorized, gold-plated copper enclosures in a single dilution refrigerator cool down.  The dilution refrigerator contained two independent measurement chains consisting of microwave attenuators, filters, 1x6 coaxial switches, isolators, directional couplers, and amplifiers (one Josephson Traveling Wave Parametric Amplifier\cite{Macklin2015} and one High Electron Mobility Transistor amplifier).  A vector network analyzer was used to measure each resonator's microwave transmission spectrum  over a range of internal circulating photon numbers $n_p$ at 25 mK in a magnetically shielded, light-tight environment.  These transmission spectra were then fitted to determine the intrinsic quality factor of the resonator in the low-power limit ($n_p \sim 1$) where $Q_i$ is mostly dominated by TLS-related losses, and the high-power limit ($n_p \sim 10^6$), where $Q_i$ is dominated by power-independent losses such as vortices\cite{Song2009}, quasiparticles\cite{Visser2011, Nsanz2014}, or radiation/package losses \cite{Sage2011}.  In order to reduce the systematic and variable losses contributed by these power-independent mechanisms and thereby determine the aggregate losses that are solely due to interface and substrate TLS, we subtracted the high-power losses from the low-power losses to determine a TLS-limited quality factor $Q_{TLS}$.   Additional details on the measurement apparatus, techniques, and data analysis can be found in Ref. \cite{Calusine2018}. \par

\begin{figure}
\includegraphics[scale=.9]{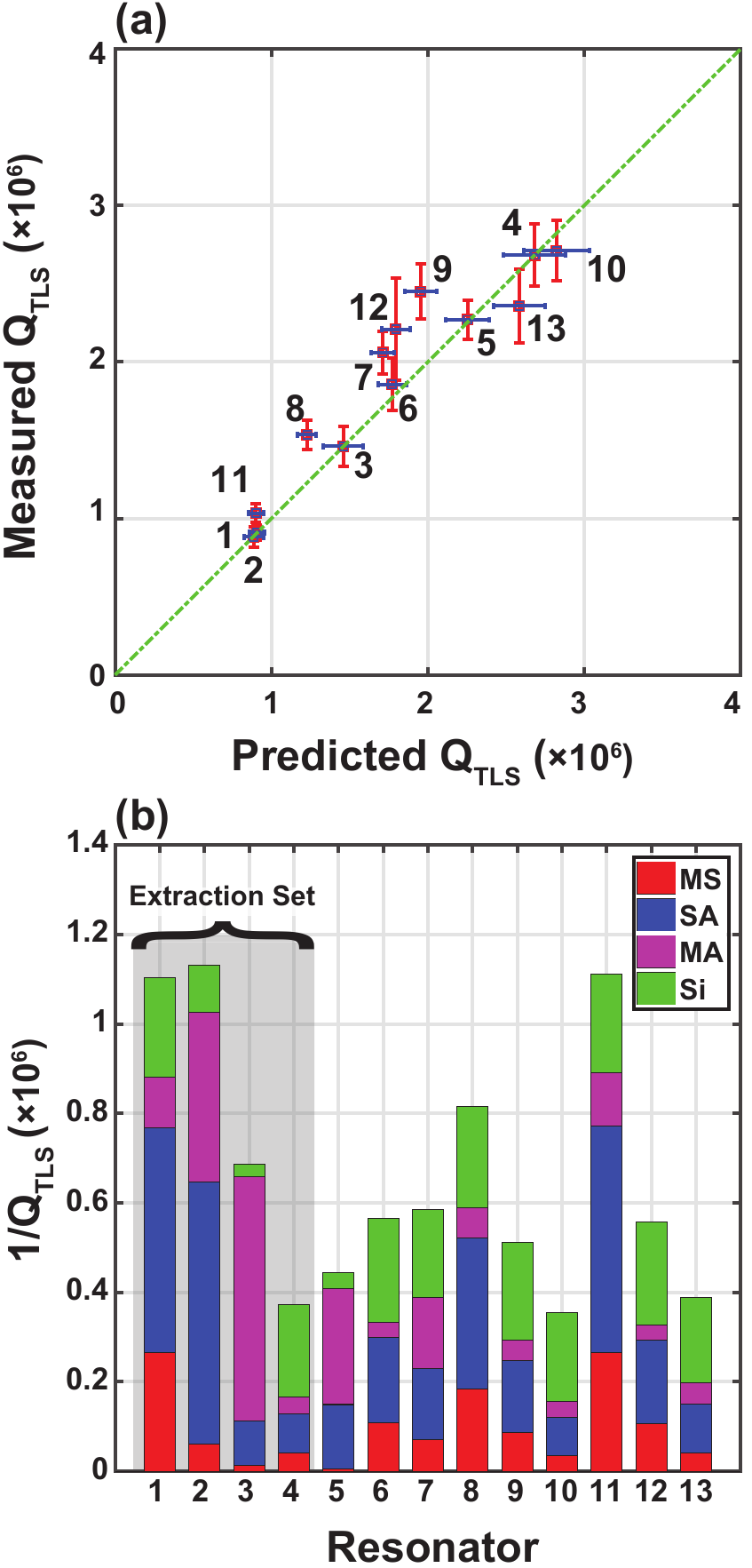}
\caption{\label{fig5} (a) Measured vs. predicted $Q_{TLS}$ and (b) corresponding loss contributions of thirteen measured resonators.}
\end{figure}

We combined the measured $Q_{TLS}$ values with the simulated participation matrix $\boldsymbol{P}$ to extract the loss factor vectors $\vec{x}$ in Eq. 1 using a linear least-squares fit.  To estimate the uncertainty in the resulting solutions, we used Monte Carlo error analysis.   Each input trial case for the Monte Carlo analysis was selected from the estimated distribution of $Q_{TLS}$ we determined by measuring approximately 30 CPW resonators for each of the 4 geometries.  The mean and standard deviation of these input distributions were determined from the mean and standard error of the measured $Q_{TLS}$ values.  A range of loss factor vectors $\vec{x}$ was then extracted using the 4x4 participation matrix $\boldsymbol{P}$ and the matrix representation in Eq. 1 from N=10000 repetitions of the Monte Carlo simulation.

For quantifying and predicting the losses that set $Q_{TLS}$, the loss factors in Eq. 1 are sufficient.  However, it is often desirable to estimate the loss tangents of the dielectric regions using reasonable assumptions for the interface dielectric thickness \textit{t} and permittivity $\epsilon$. Using values similar to those typically assumed in the literature for thicknesses and dielectric values of the TLS defect regions, e.g. Refs. \cite{Calusine2018, Wenner2011, Quintana2014, Gambetta2017},   t$_{MS}$ = 2 nm, t$_{SA}$ = 2 nm, t$_{MA}$ = 2 nm, $\epsilon_{MS}$ = 11.4$\epsilon_0$, $\epsilon_{SA}$ = 4$\epsilon_0$, $\epsilon_{MA}$ = 10$\epsilon_0$, we can ascribe loss tangents to the individual defect regions. The output histograms resulting from the Monte Carlo estimation of these loss tangents are shown in Fig. 3 and the mean values and associated 95\% confidence intervals are given in Eq. \ref{eq3}:

\begin{equation} \label{eq3}
[\tan{\delta}]_{\textrm{Range}} = \Bigg[\begin{smallmatrix} \tan{\delta}_{\textrm{MS}}\\\tan{\delta}_{\textrm{SA}}\\\tan{\delta}_ {\textrm{MA}}\\\tan{\delta}_{\textrm{Si}}\end{smallmatrix} \Bigg]=\Bigg[ \begin{smallmatrix} 4.8\times10^{-4} \pm \  2 \times10^{-4}  \\ 1.7\times10^{-3} \pm \  4 \times10^{-4}\\ 3.3\times10^{-3} \pm \  4 \times10^{-4} \\ 2.6\times10^{-7} \pm \  4 \times10^{-8}\end{smallmatrix} \Bigg]
\end{equation}

The Gaussian output distributions for each loss tangent indicate that this combination of values represent a stable, unique estimation of the losses for each dielectric region in our devices.  

In order to verify this participation-based loss model, we used the loss factors presented in Eq. 2 to predict the aggregate losses for nine additional, distinct resonator geometries with interface participation and total $Q_{TLS}$ that differ significantly from the device set used for loss factor extraction. These additional devices had center trace widths \textit{w} ranging from 6 $\mu$m to 28 $\mu$m, gaps to ground \textit{g} ranging from 1 $\mu$m to 14 $\mu$m, and trench depths \textit{d} ranging from 280 nm to 10.9 $\mu$m.  The resulting $Q_{TLS}$ ranged from $\sim 1\times10^6$ to $\sim 2.7\times10^6$.  A comparison between the predicted and measured $Q_{TLS}$ for all thirteen resonator geometries is shown in Fig. 4(a). The green dashed line in Fig. 4(a) represents the ideal case where the measured $Q_{TLS}$ is equal to the predicted $Q_{TLS}$. The red error bars represent the 95\% confidence interval for the mean measured $Q_{TLS}$ while the blue error bars show the 95\% confidence interval of the predicted $Q_{TLS}$ as determined by Monte Carlo simulations. The bar graph in Fig. 4(b) shows the absolute contributions to losses $Q_{TLS}^{-1} = \sum{Q_{i}^{-1}} , i \in \{MS,SA,MA,Si\}$ in each of the thirteen CPW resonators whose quality factors are shown in Fig. 4(a). This confirms that the four CPW resonator test cases used to extract the individual region losses each emphasize a different single dielectric while minimizing the others; the devices labeled 1-4 (the extraction set) in Fig. 4(b) and shown in Fig. 2(a-d), respectively, have the largest contribution to their aggregate losses from the MS, SA, MA, and Si dielectric layers, respectively, relative to the other devices.  \par

In order to estimate how many devices must be measured in order to obtain a unique set of loss tangents, we performed a series of simulated experiments using the loss tangents in Eq. 2 as known target values.  Furthermore, as part of this analysis, we compared the isotropic resonator designs labeled 1-4 in this work with the four anisotropic cases that produce the lowest condition number participation matrix from the device set measured in Ref. \cite{Calusine2018}.  The simulated experiment proceeded as follows: First, the target loss factors were combined with the simulated device participation matrices using Eq. 1 to determine the ideal $Q_{TLS}$.  Then, to determine the least-squares solution subject to realistic experimental uncertainty, we randomly selected a number N of  $Q_{TLS}$ values from a distribution typical of real measured values to simulate the measurement of N resonators.  These values were then used to estimate the parent $Q_{TLS}$ distribution for loss factor extraction with Monte Carlo error analysis.  This process is then repeated many times to generate metrics for comparing the extraction success.  The metric we use to compare the loss factor extraction for the two device sets is the ``worst case'' uncertainty set by the largest 95\% confidence intervals that result from any single instance of many repeated iterations of the simulated experiment. To determine the number of devices required to obtain a unique solution, we performed the extraction for different device quantities.  \par

Figure 5 shows the worst case uncertainties in the extracted loss tangents for different numbers of total devices for the isotropically and anisotropically etched cases. The results show that for both cases, measuring more devices reduces the loss factor extraction uncertainty, as expected from the resulting improvements in the estimate of $Q_{TLS}$ .  However, only for the isotropic etching case does the MS, SA, and MA loss uncertainty achieve a non-zero lower bound that indicates a stable solution within a reasonable number of measured resonators (approximately 120 resonators or two dilution refrigerator cooldowns).   When only an upper bound is assigned to the loss factor distribution, such as in MS, SA, and MA loss factors extracted from the anisotropically etched devices, the histograms of the Monte Carlo error analysis output are approximately evenly distributed throughout the uncertainty range of output solutions.  This result is similar to previous data in the literature that bounded the possible interface loss magnitudes but was unable to assign a unique solution for each device interface.  In contrast, these results show that the loss factors determined from the isotropically etched devices exhibit clear upper and lower bounds resolved for all 4 interfaces once a sufficient number of devices are characterized, consistent with what we observe experimentally. Furthermore, these results show that this technique requires only moderate measurement resources to determine the separate contributions to TLS losses in devices that are fabricated using processes similar to those used in this work.\par

\section{Discussion}

The unique estimation of the interface loss tangents shown in Fig. 3 was enabled by three aspects of this analysis.  First and foremost, we designed a set of isotropically etched CPWs to form a participation matrix $\boldsymbol{P}$ that is significantly better conditioned than is possible with planar designs or anisotropic trenching.  Second, we measured many nominally identical copies of the same device to compensate for device-to device variation.  This allowed us to generate an accurate estimate of the mean $Q_{TLS}$ associated with each geometry and to determine the $Q_{TLS}$ statistics required for estimation of the loss factor uncertainty using Monte Carlo techniques.  Finally, cross-sectional imaging of each device geometry greatly refined the accuracy of the electrostatic simulations used to determine each geometry's interface participation.\par

\begin{figure}
\includegraphics[scale=1]{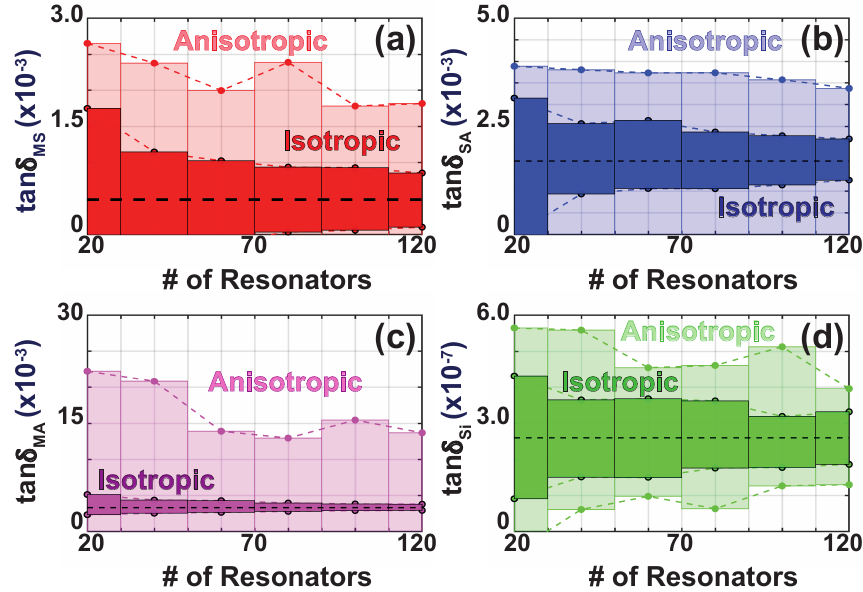}
\caption{\label{fig6} Simulated maximum uncertainty of the dielectric loss tangent extractions vs. number of measured resonators for isotropically and anisotropically etched resonators.  The black dashed lines indicates the target loss tangents used in simulations. (a) MS. (b) SA. (c) MA. (d) Si.}
\end{figure}

The good agreement between the measured and predicted $Q_{TLS}$ for devices with a wide range of total $Q_{TLS}$ and participation ratios shown in Fig. 4(a) demonstrates the accuracy and utility of the loss factor analysis.  In addition to this predictive power, this technique can be used as a diagnostic tool for assessing the relative contributions to total losses from different interfaces for a given geometry, as shown in Fig. 4 (b).  Resonator 12, for example, with  (\textit{w}, \textit{g}, \textit{d}) = (22 $\mu$m, 11 $\mu$m, 0.41 $\mu$m), has a cross section that is the most similar to many untrenched resonator and qubit capacitors used in superconducting qubit circuits.\cite{Barends2013}  For this case, Fig. 4(b) demonstrates that although the SA and silicon dielectric regions have the largest contributions to the aggregate losses, all four regions have non-negligible contributions. \par

In summary, we have combined statistical characterization of sets of specially designed, isotropically etched CPW resonators with detailed EM modeling and Monte Carlo error analysis in order to uniquely determine the individual interface losses in superconducting microwave resonators. To the best of our knowledge, this is the first time that TLS-related losses of the silicon substrate and individual interface dielectrics have been determined for a superconducting quantum circuit.  The determination of these values enables the construction of a predictive, participation-based model for aggregate device losses that we verified using a series of superconducting CPW resonators with a range of participation ratios and total $Q_{TLS}$.  This technique for distinguishing the relative contributions from individual material interface losses could be utilized to drive improvements in qubits and resonator design. Alternatively, the knowledge generated using this process can provide interface-specific feedback for improving fabrication processes or qualifying fabrication process changes.  In general, this technique stands to significantly enhance our ability to compare different materials and fabrication process for improving the performance of superconducting quantum circuits.  \par

\section{Acknowledgments}
We gratefully acknowledge M. Augeri, P. Baldo, M. Cook, R. Das, E. Dauler, M. Hellstron, V. Iaia, K. Magoon, X. Miloshi, P. Murphy, B. Osadchy, R. Slattery, C. Thoummaraj, and D. Volfson at MIT Lincoln Laboratory for technical assistance. This material is based upon work supported by the Department of Defense under Air Force Contract No. FA8721-05-C-0002 and/or FA8702-15-D-0001. Any opinions, findings, conclusions or recommendations expressed in this material are those of the authors and do not necessarily reflect the views of the Department of Defense.

\section{Author Contributions}

W.W. performed all EM simulations and device design with guidance from G.C. and A.M.  A. M. fabricated and characterized the devices with feedback from G.C., D.K., J.Y., and W.W.  G.C., E.G. and A.S. measured the device properties. G.C. and W.W. analyzed the data and performed statistical analysis.  D. R., W.D.O contributed to the experimental design and analysis.  G.C and W.W wrote the manuscript with feedback from coauthors.  

\section{Competing Interests}
The authors declare no competing financial interests.

\pagebreak
\widetext
\begin{center}
\textbf{\LARGE Supplementary Materials for ``Determining interface dielectric losses in superconducting coplanar waveguide resonators''}
\end{center}
\twocolumngrid

\setcounter{equation}{0}
\setcounter{figure}{0}
\setcounter{table}{0}
\setcounter{page}{1}
\renewcommand{\thefigure}{S\arabic{figure}}
\renewcommand{\theequation}{S\arabic{equation}}
\renewcommand{\bibnumfmt}[1]{[S#1]}
\renewcommand{\citenumfont}[1]{S#1}

\section{Participation Ratio Model}

In the main text, we apply a participation ratio-based model to determine losses in distinct device dielectric regions.  The participation ratio of a bulk material or material interface dielectric region, $p_i$, is defined in Eq. \ref{eq1}:

\begin{equation} \label{eq1}
p_i = \frac{U_i}{U_{tot}} = \frac{\int_i{\frac{\epsilon_i | E |^2}{2} }}{\int_V{\frac{\epsilon_i | E|^2}{2 }}}
\end{equation}

where $U_i$ is the electric field energy stored in region \textit{i}, $U_{tot}$ is the total electric field energy stored in all regions, $E$ is the local electric field, and $\epsilon_i$ is the permittivity of the dielectric region \textit{i}.  The volume integrals in the numerator and denominator occur over region \textit{i} and the total volume, respectively.  If the actual dielectric layer participation ratios $p_i$ were known exactly, the TLS-limited Q value, $Q_{TLS}$, for a resonator would be given by Eq. \ref{eq2}: 

\begin{equation} \label{eq2}
\frac{1}{Q_{TLS}}=\sum_i{p_i\tan{\delta_i}}
\end{equation}

where $\tan{\delta_i}$ is the loss tangent of each material interface dielectric or bulk dielectric region \textit{i}.  The calculation of participation ratios requires estimates of the permittivity and thickness of each dielectric interface region, and these values are usually not well known. As a result, loss tangents cannot be directly derived from measurements.\par

Previous studies have typically assumed a value for one or both of the interface region thickness and permittivity parameters when calculating participation ratios. The material interface dielectric regions are thin as compared to other device dimensions, and their participation ratios scale with thickness and dielectric value \cite{sWenner2011}. As result, we can alternatively use `loss factors' $x_i$ in place of loss tangents, as defined in Eq. \ref{eq3} and \ref{eq4}:

\begin{equation} \label{eq3}
x_{i,\parallel} = \frac{(t_i/t_{nom,i})}{(\epsilon_{nom,i}/\epsilon_i)}\tan{\delta_i}
\end{equation}

\begin{equation} \label{eq4}
x_{i,\bot} = \frac{(t_i/t_{nom,i})}{(\epsilon_i/\epsilon_{nom,i})}\tan{\delta_i}
\end{equation}

Equation \ref{eq3} defines loss factors for the case where the electric field is parallel to the interface dielectric region, and Eq. \ref{eq4} defines loss factors for where the electric field is orthogonal to the interface dielectric region. The loss factors are dimensionless and account for the loss tangents and scaling of the actual defect layer thicknesses and permittivities, $t_i$ and $\epsilon_i$, relative to those used in the COMSOL participation ratio simulations, $t_{nom,i}$ and $\epsilon_{nom,i}$. Note that Eqs.  \ref{eq3} and  \ref{eq4} apply only to the thin material interface dielectric regions. The large, bulk substrate (in our case, silicon), is a special case because its participation ratio is not very sensitive to tiny, nanometer size inaccuracies in material interface region thicknesses and its dielectric value is fairly well known. As a result, we can make the further approximation that, for the case of the bulk substrate, the loss factor equals the loss tangent:

\begin{equation} \label{eq200}
x_{i,sub} \approx \tan{\delta_{i,sub}}
\end{equation}

Equation \ref{eq2} can then be expressed as a function of simulated participation ratios $P_i$ and loss factors as shown in Eq. \ref{eq5}:

\begin{equation} \label{eq5}
\frac{1}{Q_{TLS}}=\sum_i{P_ix_i}
\end{equation}

We partition the bulk and interface dielectrics layers of the CPW structures into four regions: metal-to-substrate (MS), substrate-to-air/vacuum (SA), metal-to-air/vacuum (MA), and the silicon substrate (Si) as described the main text. The following assumptions were applied for analyzing loss contributions from the thin interface dielectric regions:

\begin{equation} \label{eq7}
P_{MS}\approx P_{MS,\bot} \gg P_{MS,\parallel}
\end{equation}
\begin{equation} \label{eq8}
P_{MA}\approx P_{MA,\bot} \gg P_{MA,\parallel}
\end{equation}
\begin{equation} \label{eq9}
P_{SA}\approx P_{SA,\parallel} \gg P_{SA,\bot}
\end{equation}

Equations \ref{eq7} and \ref{eq8} result from the boundary conditions imposed by assuming perfectly superconducting metals.  Equation \ref{eq9} was verified empirically using finite-element simulations of the CPW resonator cross-sections performed in COMSOL.  This assumption simplifies loss factor analysis by limiting the solution outcome to four loss factors instead of five. \par

The relations in Eq. \ref{eq200} and Eqs. \ref{eq7}-\ref{eq9} result in the following approximations for the loss factors:
\begin{equation} \label{eq10}
x_{MS}\approx x_{MS,\bot}
\end{equation}
\begin{equation} \label{eq11}
x_{MA}\approx x_{MA,\bot}
\end{equation}
\begin{equation} \label{eq12}
x_{SA}\approx x_{SA,\parallel}
\end{equation}
\begin{equation} \label{eq13}
x_{Si}\approx \tan{\delta_{Si}}
\end{equation}
$Q_{TLS}$ can then be written as a sum of individual dielectric region components as shown in Eq. \ref{eq14}.

\begin{equation} \label{eq14}
\frac{1}{Q_{TLS}}=\frac{1}{Q_{MS}}+\frac{1}{Q_{SA}}+\frac{1}{Q_{MA}}+\frac{1}{Q_{Si}}
\end{equation}

Using Eqs. \ref{eq7}-\ref{eq13}, Eq. \ref{eq14} can be rewritten as:

\begin{equation} \label{eq15}
\frac{1}{Q_{TLS}}=P_{MS}x_{MS}+P_{SA}x_{SA}+P_{MA}x_{MA}+P_{Si}x_{Si}
\end{equation}

Loss factors can be obtained directly by solving a system of equations of the form of Eq. \ref{eq15} for a set of CPW resonators with simulated bulk and interface dielectric layer participation ratios and the measured $Q_{TLS}$.  This system of equations is represented in matrix form in Eq. \ref{eq6}:

\begin{equation} \label{eq6}
\Big[\frac{\bf 1}{\bf Q_{TLS}}\Big]=[\bf P][X]
\end{equation}

where [$\frac{\bf 1}{\bf Q_{TLS}}$] is a column vector where the number of rows equals the number of unique CPW resonator geometries, [\textit{\bf P}] is the participation ratio matrix where the number of rows equals the number of unique CPW resonator geometries and the number of columns equals the number of relevant interface or bulk dielectric regions, and [\textbf{X}] is a column vector where the number of rows equals the number of relevant interface or bulk dielectric regions.\par

\section{Participation ratio proportionality vs. trench depth for isotropic trenching}

Finite element simulations of device participation ratios performed in this and other works have demonstrated that isotropic trenching of CPW resonators and planar qubits breaks the approximately proportional scaling of the MS and SA participation ratios vs. trench depth exhibited by anisotropically trenched devices.\cite{sCalusine2018,sWang2015} Fig. 1(b) in the main text shows a cross-section of an isotropically trench CPW resonator depicting the trench depth, \textit{d}, and the MS, SA, MA, and Si dielectric regions. Table \ref{tab1} shows the ratio of the simulated MS participation ratio to the SA participation ratio vs. trench depth for an isotropically trenched CPW resonator with a 6 $\mu$m center trace width and 3 $\mu$m gap-to-ground.  This trend demonstrates that the MS participation ratio decreases significantly faster than the SA participation ratio as a function of trench depth. This difference in scaling of participation ratios vs. trench depth between isotropically and anisotropically trenched resonators leads to participation ratio matrices with lower condition numbers for the isotropically trenched case.  This, in turn, results in lower uncertainty for extracted loss factors or loss tangents.

\begin{table}
\begin{tabular}{||c c||} 
\hline
d($\mu$m) & MS part./SA part. \\
\hline \hline
0.13 &  0.81 \\
\hline
0.72 & 0.68 \\
\hline
1.31 & 0.59 \\
\hline
1.91 & 0.49 \\
\hline
2.50 & 0.34 \\
\hline
\end{tabular}
\caption{Ratio of simulated MS participation ratio to the SA participation ratio vs. trench depth, d, for an isotropically trench CPW resonator with (w, g) = (6 $\mu$m, 3 $\mu$m).}
\label{tab1}
\end{table}

\section{Participation ratio matrix condition number}

\begin{figure*} 
\includegraphics[scale=1]{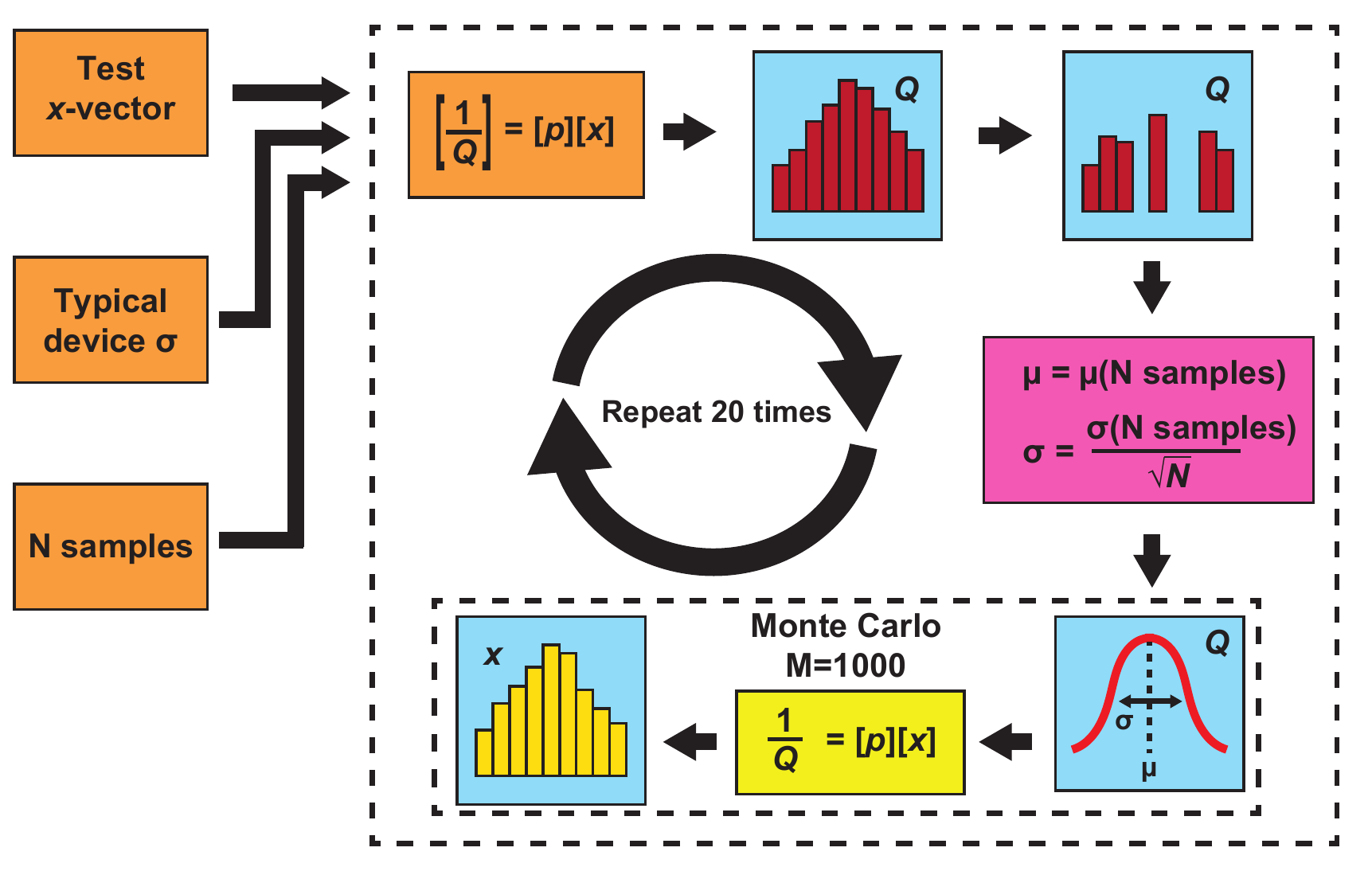}
\caption{Flow chart of worst-case loss factor extraction uncertainty simulation.}
\label{FigS7}
\end{figure*}

To extract the loss factors $x_i$ for each dielectric region for a set of device geometries and their measured $Q_{TLS}$, one must solve an equation of the form of Eq. \ref{eq6}.  When inverting the participation ratio matrix equation in Eq. \ref{eq6} to solve for [\textbf{X}], the condition number of the participation ratio matrix, $\kappa(\bf P)$, determines how sensitive the loss factor solutions are to the variation in the measured inputs, $\frac{1}{Q_{TLS}}$.  In general, when solving sets of linear equations, a large condition number generates unstable solutions with large confidence intervals for even small amounts of input uncertainty. \par
 
In an idealized experiment to extract the TLS-related losses from four separate dielectric regions, in principle, four CPW resonators could be designed such that each design was sensitive to the losses in only a single dielectric region. For the case of the four dielectric regions presented in the main text, these four distinct device designs would each have a 100$\%$ participation ratio for one of the MS, SA, MA, or Si dielectric regions.  The participation ratio matrix for this set of set of CPW resonators is the 4x4 identity matrix shown in  Eq. \ref{eq100}: 

 \begin{equation}  \label{eq100}
{\bf P_{ideal}} = 
\begin{bmatrix} 1&0&0&0\\0&1&0&0\\0&0&1&0\\0&0&0&1 \end{bmatrix}
\end{equation}

The condition number, $\kappa$, of this participation ratio matrix is unity: 

 \begin{equation}  \label{eq101}
{\kappa}{(\bf P_{ideal})} = 1
\end{equation}

In this case, the uncertainties of the extracted loss factors can be determined trivially: the fractional uncertainties of the solutions are equal to the fractional uncertainties of the $\frac{1}{Q_{TLS}}$: 

 \begin{equation}  \label{eq102}
\frac{\sigma (X)}{\mu (X)} = \frac{\sigma ( \frac{1}{Q_{TLS}})}{\mu ( \frac{1}{Q_{TLS}})}
\end{equation}

where $\sigma$  and $\mu$ are the standard deviations and means, respectively.

However, for realistic CPW resonator devices that are either purely planar or anisotropically trenched, the resulting participation ratio matrix is drastically different. In general, the participation ratio matrix rows will exhibit a greater degree of collinearity, resulting in a much more singular matrix. Equation \ref{eq103} shows an example participation ratio matrix for four devices from Ref. \onlinecite{sCalusine2018} chosen to exhibit minimally collinear participation ratio vectors:

\begin{equation}  \label{eq103}
{\bf P_{ani}} = ( \frac{1}{100} )
\begin{bmatrix} 0.1472&0.0708&0.0042&90.9972\\0.2680&0.2161&0.0264&70.0429\\0.2104&0.1304&0.0102&82.5398\\0.0878&0.0466&0.0027&87.6758 \end{bmatrix}
\end{equation}

The participation ratios shown in Eq. \ref{eq103} were calculated from simulations of CPW cross-sections performed in COMSOL. The simulations were performed using the following dielectric region permittivities and thicknesses: $\epsilon_{MS}$=$\epsilon_{Si}$=11.35$\epsilon_0$, $\epsilon_{SA}$=4$\epsilon_0$, $\epsilon_{Si}$=10$\epsilon_0$, $th_{MS}$=$th_{SA}$=$th_{SA}$=10 nm.

The condition number of this participation ratio matrix is significantly higher than the idealized case: 

\begin{table*}
\begin{tabular}{||c c c c c c c||} 
\hline
$tan{\delta}$ & This work & Ref. \onlinecite{sOConnell2008} & Ref. \onlinecite{sQuintana2014} & Ref. \onlinecite{sGambetta2017} & Ref. \onlinecite{sWenner2011} & Ref. \onlinecite{Wang2015}\\
\hline \hline
$\tan{\delta_{MS}} $& $4.8\times10^{-4}$ & - & - & - & - & $ <2.6\times10^{-3}$\\
\hline
$\tan{\delta_{SA}} $& $1.7\times10^{-3}$& $3.1\times10^{-4}$ & - & - & - & $ <2.2\times10^{-3}$\\
\hline
$\tan{\delta_{MA}}$ & $3.3\times10^{-3}$& $1.5\times10^{-3}$ & $2.6\times10^{-3}$ & - & $2\times10^{-3}$ & $2.1\times10^{-2}$\\
\hline
$\tan{\delta_{Si}}$ & $2.6\times10^{-7}$& - & - & $<5\times10^{-7}$& - & $<1\times10^{-6}$\\
\hline
\end{tabular}
\caption{Comparisons between reported interface and bulk dielectric loss tangents.}
\label{tabS2}
\end{table*}

 \begin{equation}  \label{eq104}
{\kappa}{(\bf P_{ani})} = 110,201 
\end{equation}

The large condition number of this participation ratio matrix results in loss factor solutions with large error bars relative to the input uncertainty:

 \begin{equation}  \label{eq105}
\frac{\sigma (X)}{\mu (X)} \gg \frac{\sigma ( \frac{1}{Q_{TLS}})}{\mu ( \frac{1}{Q_{TLS}})}
\end{equation}

In contrast, specially designed, isotropically trenched CPW resonators instead generate participation ratio matrices with rows that are less collinear, resulting in a less singular matrix with a smaller condition number.  For example, the participation ratio matrix in Eq. \ref{eq106} results from the four designs used to extract loss factors in the main text:

 \begin{equation}  \label{eq106}
{\bf P_{iso}} = ( \frac{1}{100} )
\begin{bmatrix} 0.2738&0.1473&0.0174&86.1487\\0.0629&0.1716&0.0580&41.0988\\0.0137&0.0290&0.0837&10.9642\\0.0416&0.0259&0.0056&80.5158 \end{bmatrix}
\end{equation}

The condition number of this matrix is: 

 \begin{equation}  \label{eq107}
{\kappa}{(\bf P_{iso})} = 2,001
\end{equation}

which is approximately 55 times smaller than for the anisotropic case presented above.  As a result, this case generates stable loss factor solutions with relatively small uncertainty such as those we present in the main text. \par 

\section{Simulated loss factor extraction experiment}

For a given set of resonator designs and expected loss factors, one can combine participation ratio calculations with realistic values for device variance to perform `simulated experiments' in order to estimate the resource requirements for successful loss factor extraction.  In the main text, we describe using this technique to estimate the necessary number of resonator measurements for successful extraction of the known loss factors presented in this work.   Fig. \ref{FigS7} shows a block diagram of the MATLAB algorithm used to simulate the worst-case loss factor extraction uncertainty given a known loss factor vector as input.  First, a known loss-factor vector is assumed as input (orange block on upper left). Next, the participation ratio matrix of the test set of resonators is used to compute the $Q_{TLS}$ values associated with the resonators.  We then generate a standard normal distribution of $Q_{TLS}$ using the calculated value for the distribution mean and a standard deviation consistent with values typically observed in experiment.  To simulate the experimental process of characterizing device statistics by randomly sampling from an unknown parent distribution, a set of N samples for $Q_{TLS}$ are then randomly chosen from these distributions for each resonator design. Next, just as done in our experiments, a mean value and standard deviation value is estimated from the N sampled $Q_{TLS}$  values (pink block in Fig. \ref{FigS7}). Finally, the loss factor extraction is performed using linear least squares estimation with Monte Carlo error analysis (N=1000) (yellow block in Fig. \ref{FigS7}). This results in a mean solution and output uncertainty for each loss factor.  We define the `worst case error' as the most extreme upper and lower bounds of the 95\% confidence interval of each set of extracted loss factors over 20 repetitions of the simulated loss factor extraction experiment.  \\ \par

\section{Comparison to previous work}

Table \ref{tabS2} shows the comparisons between the best-fit loss tangents obtained in this work and the loss tangents of comparable material interfaces studied in the literature. For all interfaces, the loss tangents extracted in this work are comparable to previous estimates or fall within the bounds determined in those experiments.   In comparing these values, some inconsistency should be expected resulting from the ambiguity in the expected dielectric constants and interface thicknesses.  \par

\end{document}